\input harvmac
\input amssym
\input epsf

\def\unit{\relax{\rm 1\kern-.26em I}}
\def\nada{\relax{\rm 0\kern-.30em l}}
\def\tilde{\widetilde}

% \draftmode

%\def\Omega{\rho,\sigma,\nu  }

%% MACROS
\noblackbox
\def\IL{\relax{\rm I\kern-.18em L}}
\def\IH{\relax{\rm I\kern-.18em H}}
\def\IR{\relax{\rm I\kern-.18em R}}
\def\IC{\relax\hbox{$\inbar\kern-.3em{\rm C}$}}
\def\IZ{\relax\ifmmode\mathchoice
{\hbox{\cmss Z\kern-.4em Z}}{\hbox{\cmss Z\kern-.4em Z}}
{\lower.9pt\hbox{\cmsss Z\kern-.4em Z}} {\lower1.2pt\hbox{\cmsss
Z\kern-.4em Z}}\else{\cmss Z\kern-.4em Z}\fi}

\def\CJ {{\cal J}}

\def\CL {{\cal L}}
\def\CV {{\cal V}}
\def\CO {{\cal O}}

\def\CA{{\cal A}}

%% MORE MACROS

\def\CO {{\cal O}}

\def\CV{{\cal V }}

\def\Tr{{\rm Tr}}

\font\manual=manfnt \def\dbend{\lower3.5pt\hbox{\manual\char127}}

\def\IZ{\relax\ifmmode\mathchoice
{\hbox{\cmss Z\kern-.4em Z}}{\hbox{\cmss Z\kern-.4em Z}}
{\lower.9pt\hbox{\cmsss Z\kern-.4em Z}} {\lower1.2pt\hbox{\cmsss
Z\kern-.4em Z}}\else{\cmss Z\kern-.4em Z}\fi}
\def\half {{1\over 2}}

\def\lfm#1{\medskip\noindent\item{#1}}

\def\bar{\overline}

\def\rt2{\sqrt{2}}
\def\irt2{{1\over\sqrt{2}}}

%  \slashchar puts a slash through a character to represent contraction
%  with Dirac matrices. Use \not instead for negation of relations, and use
%  \hbar for hbar.
\def\slashchar#1{\setbox0=\hbox{$#1$}           % set a box for #1
   \dimen0=\wd0                                 % and get its size
   \setbox1=\hbox{/} \dimen1=\wd1               % get size of /
   \ifdim\dimen0>\dimen1                        % #1 is bigger
      \rlap{\hbox to \dimen0{\hfil/\hfil}}      % so center / in box
      #1                                        % and print #1
   \else                                        % / is bigger
      \rlap{\hbox to \dimen1{\hfil$#1$\hfil}}   % so center #1
      /                                         % and print /
   \fi}

\def\foursqr#1#2{{\vcenter{\vbox{
    \hrule height.#2pt
    \hbox{\vrule width.#2pt height#1pt \kern#1pt
    \vrule width.#2pt}
    \hrule height.#2pt
    \hrule height.#2pt
    \hbox{\vrule width.#2pt height#1pt \kern#1pt
    \vrule width.#2pt}
    \hrule height.#2pt
        \hrule height.#2pt
    \hbox{\vrule width.#2pt height#1pt \kern#1pt
    \vrule width.#2pt}
    \hrule height.#2pt
        \hrule height.#2pt
    \hbox{\vrule width.#2pt height#1pt \kern#1pt
    \vrule width.#2pt}
    \hrule height.#2pt}}}}
\def\psqr#1#2{{\vcenter{\vbox{\hrule height.#2pt
    \hbox{\vrule width.#2pt height#1pt \kern#1pt
    \vrule width.#2pt}
    \hrule height.#2pt \hrule height.#2pt
    \hbox{\vrule width.#2pt height#1pt \kern#1pt
    \vrule width.#2pt}
    \hrule height.#2pt}}}}
\def\sqr#1#2{{\vcenter{\vbox{\hrule height.#2pt
    \hbox{\vrule width.#2pt height#1pt \kern#1pt
    \vrule width.#2pt}
    \hrule height.#2pt}}}}
\def\square{\mathchoice\sqr65\sqr65\sqr{2.1}3\sqr{1.5}3}

\def\figin{\epsfcheck\figin}\def\figins{\epsfcheck\figins}
\def\epsfcheck{\ifx\epsfbox\UnDeFiNeD
\message{(NO epsf.tex, FIGURES WILL BE IGNORED)}
\gdef\figin##1{\vskip2in}\gdef\figins##1{\hskip.5in}% blank space instead
\else\message{(FIGURES WILL BE INCLUDED)}%
\gdef\figin##1{##1}\gdef\figins##1{##1}\fi}
\def\DefWarn#1{}
\def\figinsert{\goodbreak\midinsert}
\def\ifig#1#2#3{\DefWarn#1\xdef#1{fig.~\the\figno}
\writedef{#1\leftbracket fig.\noexpand~\the\figno}%
\figinsert\figin{\centerline{#3}}\medskip\centerline{\vbox{\baselineskip12pt
\advance\hsize by -1truein\noindent\footnotefont{\bf
Fig.~\the\figno:\ } \it#2}}
\bigskip\endinsert\global\advance\figno by1}

\lref\gmreview{
  G.~F.~Giudice and R.~Rattazzi,
  ``Theories with gauge-mediated supersymmetry breaking,''
  Phys.\ Rept.\  {\bf 322}, 419 (1999)
  [arXiv:hep-ph/9801271].
  %%CITATION = PRPLC,322,419;%%
}

\lref\BrignoleCM{
  A.~Brignole, J.~A.~Casas, J.~R.~Espinosa and I.~Navarro,
  ``Low-scale supersymmetry breaking: Effective description, electroweak
  breaking and phenomenology,''
  Nucl.\ Phys.\  B {\bf 666}, 105 (2003)
  [arXiv:hep-ph/0301121].
  %%CITATION = NUPHA,B666,105;%%
  }

%\AffleckXZ
\lref\AffleckXZ{
  I.~Affleck, M.~Dine and N.~Seiberg,
  ``Dynamical Supersymmetry Breaking In Four-Dimensions And Its
  Phenomenological Implications,''
  Nucl.\ Phys.\  B {\bf 256}, 557 (1985).
  %%CITATION = NUPHA,B256,557;%%
}

%\DineYW
\lref\DineYW{
  M.~Dine and A.~E.~Nelson,
  ``Dynamical supersymmetry breaking at low-energies,''
  Phys.\ Rev.\  D {\bf 48}, 1277 (1993)
  [arXiv:hep-ph/9303230].
  %%CITATION = PHRVA,D48,1277;%%
}

%\DineVC
\lref\DineVC{
  M.~Dine, A.~E.~Nelson and Y.~Shirman,
  ``Low-Energy Dynamical Supersymmetry Breaking Simplified,''
  Phys.\ Rev.\  D {\bf 51}, 1362 (1995)
  [arXiv:hep-ph/9408384].
  %%CITATION = PHRVA,D51,1362;%%
}

%\LutyFK
\lref\LutyFK{
  M.~A.~Luty,
  ``Naive dimensional analysis and supersymmetry,''
  Phys.\ Rev.\  D {\bf 57}, 1531 (1998)
  [arXiv:hep-ph/9706235].
  %%CITATION = PHRVA,D57,1531;%%
}

%\DineAG
\lref\DineAG{
  M.~Dine, A.~E.~Nelson, Y.~Nir and Y.~Shirman,
  ``New tools for low-energy dynamical supersymmetry breaking,''
  Phys.\ Rev.\  D {\bf 53}, 2658 (1996)
  [arXiv:hep-ph/9507378].
  %%CITATION = PHRVA,D53,2658;%%
}

%\WittenKV
\lref\WittenKV{
  E.~Witten,
  ``Mass Hierarchies In Supersymmetric Theories,''
  Phys.\ Lett.\  B {\bf 105}, 267 (1981).
  %%CITATION = PHLTA,B105,267;%%
}

%\BanksMG
\lref\BanksMG{
  T.~Banks and V.~Kaplunovsky,
  ``Nosonomy Of An Upside Down Hierarchy Model. 1,''
  Nucl.\ Phys.\  B {\bf 211}, 529 (1983).
  %%CITATION = NUPHA,B211,529;%%
}
%\KaplunovskyYX
\lref\KaplunovskyYX{
  V.~Kaplunovsky,
  ``Nosonomy Of An Upside Down Hierarchy Model. 2,''
  Nucl.\ Phys.\  B {\bf 233}, 336 (1984).
  %%CITATION = NUPHA,B233,336;%%
}

%\DimopoulosGM
\lref\DimopoulosGM{
  S.~Dimopoulos and S.~Raby,
  ``Geometric Hierarchy,''
  Nucl.\ Phys.\  B {\bf 219}, 479 (1983).
  %%CITATION = NUPHA,B219,479;%%
}

%\DermisekQJ
\lref\DermisekQJ{
  R.~Dermisek, H.~D.~Kim and I.~W.~Kim,
  ``Mediation of supersymmetry breaking in gauge messenger models,''
  JHEP {\bf 0610}, 001 (2006)
  [arXiv:hep-ph/0607169].
  %%CITATION = JHEPA,0610,001;%%
}

%\DineGU
\lref\DineGU{
  M.~Dine and W.~Fischler,
  ``A Phenomenological Model Of Particle Physics Based On Supersymmetry,''
  Phys.\ Lett.\  B {\bf 110}, 227 (1982).
  %%CITATION = PHLTA,B110,227;%%
}

%\NappiHM
\lref\NappiHM{
  C.~R.~Nappi and B.~A.~Ovrut,
  ``Supersymmetric Extension Of The SU(3) X SU(2) X U(1) Model,''
  Phys.\ Lett.\  B {\bf 113}, 175 (1982).
  %%CITATION = PHLTA,B113,175;%%
}

%\DineZB
\lref\DineZB{
  M.~Dine and W.~Fischler,
  ``A Supersymmetric Gut,''
  Nucl.\ Phys.\  B {\bf 204}, 346 (1982).
  %%CITATION = NUPHA,B204,346;%%
}

%\AlvarezGaumeWY
\lref\AlvarezGaumeWY{
  L.~Alvarez-Gaume, M.~Claudson and M.~B.~Wise,
  ``Low-Energy Supersymmetry,''
  Nucl.\ Phys.\  B {\bf 207}, 96 (1982).
  %%CITATION = NUPHA,B207,96;%%
}

\lref\tobenomura{
  Y.~Nomura and K.~Tobe,
  ``Phenomenological aspects of a direct-transmission model of dynamical
  supersymmetry breaking with the gravitino mass m(3/2) $<$ 1-keV,''
  Phys.\ Rev.\  D {\bf 58}, 055002 (1998)
  [arXiv:hep-ph/9708377].
}

\lref\IzawaGS{
  K.~I.~Izawa, Y.~Nomura, K.~Tobe and T.~Yanagida,
  ``Direct-transmission models of dynamical supersymmetry breaking,''
  Phys.\ Rev.\  D {\bf 56}, 2886 (1997)
  [arXiv:hep-ph/9705228].
  %%CITATION = PHRVA,D56,2886;%%
}

%\CheungES
\lref\CheungES{
  C.~Cheung, A.~L.~Fitzpatrick and D.~Shih,
  ``(Extra)Ordinary Gauge Mediation,''
  arXiv:0710.3585 [hep-ph].
  %%CITATION = ARXIV:0710.3585;%%
}

\lref\hiddenren{
  A.~G.~Cohen, T.~S.~Roy and M.~Schmaltz,
  %``Hidden sector renormalization of MSSM scalar masses,''
  JHEP {\bf 0702}, 027 (2007)
  [arXiv:hep-ph/0612100].
  %%CITATION = JHEPA,0702,027;%%
}

\lref\dimgiud{
 S.~Dimopoulos and G.~F.~Giudice,
  ``Multi-messenger theories of gauge-mediated supersymmetry breaking,''
  Phys.\ Lett.\  B {\bf 393}, 72 (1997)
  [arXiv:hep-ph/9609344].
  %%CITATION = PHLTA,B393,72;%%
}

\lref\pierre{S.~P.~Martin and P.~Ramond,
  ``Sparticle spectrum constraints,''
  Phys.\ Rev.\  D {\bf 48}, 5365 (1993)
  [arXiv:hep-ph/9306314].
  %%CITATION = PHRVA,D48,5365;%%
}

\lref\faraggi{
  A.~E.~Faraggi, J.~S.~Hagelin, S.~Kelley and D.~V.~Nanopoulos,
  ``Sparticle Spectroscopy,''
  Phys.\ Rev.\  D {\bf 45}, 3272 (1992).
  %%CITATION = PHRVA,D45,3272;%%
}

\lref\kawamura{
  Y.~Kawamura, H.~Murayama and M.~Yamaguchi,
  ``Probing symmetry breaking pattern using sfermion masses,''
  Phys.\ Lett.\  B {\bf 324}, 52 (1994)
  [arXiv:hep-ph/9402254].
  %%CITATION = PHLTA,B324,52;%%
}

%\MartinZB
\lref\MartinZB{
  S.~P.~Martin,
  ``Generalized messengers of supersymmetry breaking and the sparticle mass
  spectrum,''
  Phys.\ Rev.\  D {\bf 55}, 3177 (1997)
  [arXiv:hep-ph/9608224].
  %%CITATION = PHRVA,D55,3177;%%
}

\lref\spectroscopy{
 S.~Dimopoulos, S.~D.~Thomas and J.~D.~Wells,
  ``Sparticle spectroscopy and electroweak symmetry breaking with
  gauge-mediated supersymmetry breaking,''
  Nucl.\ Phys.\  B {\bf 488}, 39 (1997)
  [arXiv:hep-ph/9609434].
  %%CITATION = NUPHA,B488,39;%%
}

\lref\polchinski{ J.~Polchinski and L.~Susskind, ``Breaking Of
Supersymmetry At Intermediate-Energy,''
  Phys.\ Rev.\  D {\bf 26}, 3661 (1982).
  %%CITATION = PHRVA,D26,3661;%%
 }

\lref\dgp{ G.~R.~Dvali, G.~F.~Giudice and A.~Pomarol,
  ``The $\mu$-Problem in Theories with Gauge-Mediated Supersymmetry Breaking,''
  Nucl.\ Phys.\  B {\bf 478}, 31 (1996)
  [arXiv:hep-ph/9603238].
  %%CITATION = NUPHA,B478,31;%%
}

\lref\martinmu{ T.~S.~Roy and M.~Schmaltz, ``A hidden solution to
the $\mu/B_\mu$ problem in gauge mediation,''
  arXiv:0708.3593 [hep-ph].
  %%CITATION = ARXIV:0708.3593;%%
}

\lref\hidrentwo{
  H.~Murayama, Y.~Nomura and D.~Poland,
  ``More Visible Effects of the Hidden Sector,''
  arXiv:0709.0775 [hep-ph].
  %%CITATION = ARXIV:0709.0775;%%
}

%\DineXI
\lref\DineXI{
  M.~Dine, N.~Seiberg and S.~Thomas,
  ``Higgs Physics as a Window Beyond the MSSM (BMSSM),''
  Phys.\ Rev.\  D {\bf 76}, 095004 (2007)
  [arXiv:0707.0005 [hep-ph]].
  %%CITATION = PHRVA,D76,095004;%%
}

%\WessCP
\lref\WessCP{
  J.~Wess and J.~Bagger,
  ``Supersymmetry and supergravity,''
%\href{http://www.slac.stanford.edu/spires/find/hep/www?irn=5426545}{SPIRES entry}
{\it  Princeton, USA: Univ. Pr. (1992) 259 p}
}

%\MartinNS
\lref\MartinNS{
  S.~P.~Martin,
  ``A supersymmetry primer,''
  arXiv:hep-ph/9709356.
  %%CITATION = HEP-PH/9709356;%%
}

%%%%%%%%%%%%%%%%%%%%%%%%

\newbox\tmpbox\setbox\tmpbox\hbox{\abstractfont }
\Title{\vbox{\baselineskip12pt }} {\vbox{\centerline{
General Gauge Mediation}}}
\smallskip
\centerline{Patrick Meade, Nathan Seiberg, and David Shih}
\smallskip
\bigskip
\centerline{{\it School of Natural Sciences, Institute for
Advanced Study, Princeton, NJ 08540 USA}}
\bigskip
\vskip 1cm

\noindent
 We give a general definition of gauge mediated supersymmetry
breaking which encompasses all the known gauge mediation models. In
particular, it includes both models with messengers as well as
direct mediation models. A formalism for computing the soft terms in
the generic model is presented. Such a formalism is necessary in
strongly-coupled direct mediation models where perturbation theory
cannot be used.  It allows us to identify features of the entire
class of gauge mediation models and to distinguish them from
specific signatures of various subclasses.

\bigskip

\Date{January 2008}

\newsec{Introduction}

Gauge mediation is one of the oldest, simplest, and most robust
ways of transmitting SUSY breaking to the MSSM.  It has a number
of virtues, for instance guaranteeing flavor universality among
the MSSM sfermion masses (thus solving the SUSY flavor problem).
Unfortunately, even with the inherent simplicity of gauge
mediation there is a veritable cornucopia of models (for a review
of various types of gauge mediated models and some of the early
history see \gmreview). These models have a wide variety of
features, and often it is unclear which features are model
specific and which are generic to gauge mediation. Furthermore,
despite this long list of models, it is not obvious that all the
possibilities of gauge mediation have been completely mapped out.
For instance, as was originally envisioned \AffleckXZ, direct
mediation models can be strongly coupled.  Such models have not
yet been extensively studied, in part because there is currently
no developed framework for calculating the MSSM soft masses (for
an early work see \LutyFK).

In this paper we wish to address these points by presenting a
unified framework to describe the effects of a completely
arbitrary hidden sector. At the heart of this framework is a
careful definition of the gauge mediation mechanism itself: {\it
in the limit that the MSSM gauge couplings $\alpha_i \rightarrow
0$, the theory decouples into the MSSM and a separate hidden
sector that breaks SUSY}.  (In section 5 we will slightly extend
this definition to include various couplings to the Higgs field.)
Here are some examples:
 \lfm{1.} The most common paradigm of gauge mediation is to have
a set of weakly coupled messenger fields charged under the MSSM
and some supersymmetry breaking spurion field $X$
\refs{\DineYW\DineVC-\DineAG}. Such models fit our definition by
identifying the hidden sector as including both the supersymmetry
breaking sector and the messengers; together, these decouple as
$\alpha_i \rightarrow 0$. Clearly, we can accommodate any number
of messengers and $X$ fields. Also, various models with additional
gauge field messengers which have independent gauge coupling
constants (such as models with extra $U(1)'$s) can also be
accommodated by including these fields in the hidden sector. Note
however that gauge messenger models based on nontrivial embeddings
of the SM gauge group into larger groups such as $SU(5)_{GUT}$
(see e.g.\ \refs{\WittenKV\BanksMG\KaplunovskyYX-\DimopoulosGM}
and more recently \DermisekQJ) are not covered by our formalism,
because in these models the heavy gauge fields cannot be included
in an almost decoupled hidden sector.
 \lfm{2.} Models such as \refs{\DineGU\NappiHM\DineZB\AlvarezGaumeWY
\tobenomura-\IzawaGS} and more recently \CheungES\  involve a
weakly coupled supersymmetry breaking theory (i.e.\ an
O'Raifeartaigh-like model) with a global symmetry. These are
direct mediation models, where the messenger fields participate in
the supersymmetry breaking process.
 \lfm{3.} Direct mediation models which involve a strongly coupled hidden
sector (for a sample of such models, see \gmreview). Here, there
may not  even be identifiable messenger fields but the model still
lies within our definition of gauge mediation.

\medskip

Given this definition of gauge mediation which includes strongly
coupled theories, the computation of the soft terms in the MSSM can
proceed in perturbation theory in $\alpha_i$ but must include exact
information from the hidden sector.  This information is summarized
in a set of correlation functions of real linear superfields $\CJ$
representing the hidden sector contribution to the gauge currents of
the MSSM. In this framework it turns out that all the soft terms of
the MSSM are describable in terms of only a small number of current
correlation functions for any model of gauge mediation.

We do not provide a new toolkit for being able to calculate these
current correlation functions for an arbitrarily strongly coupled
theory.  Nevertheless given any model for a hidden sector these
correlation functions parameterize the answer for the
effects of the hidden sector on the MSSM.

Superficially, our treatment is reminiscent of the effects
described in \hiddenren\ which looked at the influence of hidden
sector running on the MSSM. However in \hiddenren, these effects
are described by two scales: a scale where one integrates out some
heavy messenger particles which couple the SUSY-breaking sector to
the MSSM, and a lower scale (presumably the scale of SUSY
breaking) where the rest of the hidden sector decouples. From our
perspective of gauge mediation there is no weakly coupled
description needed anywhere and there could in principle be only
one scale.

With the framework of representing all the effects of gauge
mediation in terms of current correlation functions we are able to
derive the most generic predictions for gauge mediation. These include:
 \lfm{$\bullet$} Flavor universality among the sfermion masses
 \lfm{$\bullet$} Sum rules for sfermions $\Tr\, Ym^2=0$ and $\Tr\,
(B-L)m^2=0$ (with nonzero FI term for hypercharge, these sum rules
are appropriately modified as shown in Section 4)
 \lfm{$\bullet$} Small A terms \lfm{$\bullet$} Gravitino LSP

\noindent Additionally there are several properties that can be
true in a large set of models when more assumptions are made, but
are not necessarily predictions of gauge mediation
 \lfm{$\bullet$} Gaugino mass unification
 \lfm{$\bullet$} Large hierarchies among
sfermions with different gauge quantum numbers
 \lfm{$\bullet$} A bino or stau NLSP

\bigskip
It is important to point out that our framework -- as we have
presented it so far -- does not allow for additional interactions
which could generate $\mu$ and $B\mu$ radiatively. Since
$U(1)_{PQ}$ needs to be broken to generate $\mu$/$B\mu$, it is
necessary to introduce interactions between the MSSM and the
hidden sector which remain even in the limit that the gauge
couplings are turned off. In section 5 we will present some
preliminary remarks about how one could extend our general
framework to include direct couplings between operators in the
hidden sector and the Higgs fields of the MSSM. A successful
solution to the $\mu$/$B\mu$ problem can then be characterized as
certain conditions that the correlators of these operators must
satisfy. We will save a more detailed analysis of this extended
framework for future work.

The outline of our paper is as follows. In section 2 we describe
the global currents in the hidden sector and their relevant
correlation functions. In section 3 we weakly gauge the global
symmetry and identify it with the SM gauge symmetry. We derive
explicit formulas for the MSSM soft masses in terms of the current
correlation functions. Section 4 contains our completely general
derivation of the $U(1)_Y$ and $U(1)_{B-L}$ sum rules for gauge
mediation. We also discuss various corrections to these sum rules
from the unknown $\mu$/$B\mu$ sector, MSSM RG evolution, and
electroweak symmetry breaking. Finally section 5 contains a
preliminary discussion of how to extend our formalism to include
the sector that generates $\mu$ and $B\mu$, and how to phrase the
$\mu$ problem of gauge mediation in this new language.  In the
appendix we show how the standard analysis of models with
messengers fits into our general framework.

\newsec{Currents in the Hidden Sector}

In this section, we will work out expressions for the currents  in
the hidden sector and their correlation functions. For simplicity,
we will only consider the case where a $U(1)$ is weakly gauged;
the generalization to nonabelian groups is straightforward. Our
conventions throughout will be chosen to agree with those of
\WessCP.

To begin, let us recall that the gauge current superfield
 \eqn\Jsfield{ \CJ = \CJ(x,\theta,\bar\theta) }
is a real linear
superfield defined by the current conservation conditions
 \eqn\Jcons{ \bar{D}^2 \CJ = D^2\CJ = 0 }
In components, it looks like
 \eqn\Jcomp{ \CJ = J +i\theta
j-i\bar\theta \bar j-\theta\sigma^\mu \bar\theta  j_\mu+{1\over2}
\theta\theta \bar\theta \bar\sigma^\mu\partial_\mu j-{1\over 2}
\bar\theta\bar\theta \theta\sigma^\mu\partial_\mu\bar j-{1\over 4}
 \theta\theta\bar\theta\bar\theta \square J }
with $j_\mu$ satisfying the condition
 \eqn\jmucond{\partial^\mu j_\mu=0.}

Current conservation and Lorentz invariance imply that the only
nonzero current one-point function is
 \eqn\copf{ \langle J(x)\rangle
= \zeta
 }
(Obviously, $\zeta$ vanishes when one generalizes from $U(1)$ to
nonabelian groups.) Meanwhile, the only nonzero current-current
correlators are\foot{The correlator $\langle j_\mu(x)J(0)\rangle$
can be nonzero only if the global symmetry is spontaneously
broken, a scenario we will not consider.  To see that, note that
current conservation requires it to be proportional to
$\partial_\mu x^{-2}$.  This behavior corresponds to an exchange
of the Goldstone boson of the spontaneously broken symmetry.}
 \eqn\currents{\eqalign{
 & \langle J(x) J(0)\rangle = {1\over x^4}C_0(x^2 M^2)\cr
 & \langle j_{\alpha}(x) \bar j_{\dot\alpha}(0) \rangle
 =-i\sigma^\mu_{\alpha\dot\alpha}\partial_\mu\left(
 {1\over x^4}C_{1/2}(x^2 M^2)\right)\cr
 & \langle j_\mu(x) j_\nu(0) \rangle= (\eta_{\mu\nu}\partial^2
 -\partial_\mu \partial_\nu)\left({1\over x^4}C_{1}(x^2
 M^2)\right)\cr
 & \langle j_\alpha(x)j_\beta(0)\rangle = \epsilon_{\alpha\beta}
 {1\over x^5}B_{1/2}(x^2 M^2)\cr
 }}
Here $M$ is some characteristic mass scale of the theory. The
function $B_{1/2}(x^2 M^2)$ is complex in general, but $\zeta$
and the functions $C_a(x^2 M^2)$ must be real.

If supersymmetry is unbroken, we readily obtain
 \eqn\susyunbroken{
 C_0 = C_{1/2}=C_1 ,\qquad B_{1/2}= 0.
 }

The correlators are all finite in position space, and their short
distance behavior is controlled by dimensional analysis. In
particular, the functions $C_a$, $B_{1/2}$ are regular as $x\to
0$. Their small $x$ behavior is determined by the operator product
expansion and the UV dimensions of the operators. Since the
identity operator always appears in the OPE of $\CO(x)^\dagger
\CO(0)$, the short distance behaviors of $C_a$ are
 \eqn\shortdistance{\lim_{x\to 0} C_0(x^2M^2)=
 \lim_{x\to 0} C_{1/2}(x^2M^2)=\lim_{x\to 0}
 C_1(x^2M^2) = c}
 (As we will see in the next section, when the global
symmetry is weakly gauged, $c$ essentially corresponds to the
change in the beta function of the gauge coupling.) Since our
theory spontaneously breaks supersymmetry, its short distance
behavior is supersymmetric, as in \susyunbroken, and hence the
constant $c$ is the same for all three correlators.  Since the
correlator $\langle J(x)J(0)\rangle$ must be positive,
 \eqn\cgtrzero{
 c>0.
 }
All of this is to be contrasted with the OPE $j_\alpha(x)j_\beta(0)$
which does not include the identity operator (one easy way to see
this is using the R-symmetry), so the leading singularity in the
OPE must have the form
 \eqn\jjope{ j_\alpha(x)j_\beta(0) \sim \epsilon_{\alpha\beta}
 x^{\Delta-5}\CO+\dots
 }
for some operator $\CO$ with dimension $\Delta>1$. (In fact, a
stronger inequality can be proven using supersymmetry.) Therefore,
 \eqn\Bsmallx{ \lim_{x\to 0} B_{1/2}(x^2M^2) = 0. }

The correlators in \currents\ can receive contributions from
contact terms at $x=0$.  These depend on the regularization scheme
and the precise definition of the theory.  In our case, where we
plan to gauge the global symmetry associated with this current,
the contact terms are determined.  These are easily obtained in
momentum space as follows.

The (Euclidean) Fourier transforms of \currents\ are
 \eqn\currentsmom{\eqalign{
 & \langle J(p) J(-p)\rangle = \tilde C_0(p^2/M^2;\,M/
 \Lambda)\cr
 & \langle j_{\alpha}(p) \bar j_{\dot\alpha}(-p) \rangle
 =-\sigma_{\alpha\dot\alpha}^\mu p_\mu \tilde C_{1/2}
 (p^2/M^2;\,M/\Lambda)\cr
 & \langle j_\mu(p) j_\nu(-p) \rangle= -(p^2\eta_{\mu\nu}
 -p_\mu p_\nu)\tilde C_1(p^2/M^2;\,M/\Lambda)\cr
 & \langle j_\alpha(p)j_\beta(-p)\rangle =
 \epsilon_{\alpha\beta}M \tilde B_{1/2}(p^2/M^2)
 }}
where a factor of $(2\pi)^4\delta^{(4)}(0)$ is understood, and
$\Lambda$ is a UV cutoff regulating the integrals
 \eqn\Cadef{\eqalign{ & \tilde C_a(p^2/M^2;\,M/\Lambda) =
 \int d^4x\, e^{ipx}{1\over x^4}C_a(x^2M^2)\cr
 & M\tilde B_{1/2}(p^2/M^2) = \int d^4x\,e^{ipx}{1\over x^5}
 B_{1/2}(x^2M^2).
 }}
Here the functions $\tilde C_a$ are again real, while $\tilde
B_{1/2}$ can be complex. As in \susyunbroken, if SUSY is unbroken,
$\tilde C_0=\tilde C_{1/2}=\tilde C_1$, $\tilde B_{1/2}=0$.
Because of \shortdistance, the $\Lambda$ dependence is
 \eqn\highmom{ \tilde C_a(p^2/M^2;\,M/\Lambda) = 2\pi^2 c
 \log({\Lambda/M}) + finite,}
where when supersymmetry is broken the finite part depends on $a$.
On the other hand, we have also indicated in \currentsmom\ that
$\tilde B_{1/2}$ is cutoff independent; this immediately follows
from \Bsmallx.

In writing \currentsmom\ we have assumed a specific choice of
contact terms at $x=0$.  They are set such that the currents
satisfy the conservation equations in momentum space. This choice
is motivated by our intention to gauge this global symmetry.  More
specifically, the contact terms which are proportional to delta
functions in coordinate space are polynomial in the momentum which
are arranged such that the form of \currentsmom\ is valid.

\newsec{Gaugino and Sfermion Masses}

Now let us weakly gauge the global symmetry of the previous
section, by coupling the currents to a vector superfield,
 \eqn\LJV{
\CL_{int} =2g \int d^4\theta \CJ \CV + \dots = g( J D -\lambda j
-\bar\lambda\bar j-j^\mu V_\mu) + \dots
 }
where we have used the Wess-Zumino gauge.  The ellipses in \LJV\
represent $\CO(g^2)$ terms including two gauge fields.  Such terms
are necessary for gauge invariance.

Our hidden sector can be strongly coupled, in which case we cannot
use perturbation theory.  However, we can still assume that $g\ll
1$ and expand the functional integral to second order in $g$. The
hidden sector contribution is captured by the two point functions
\currents\ or \currentsmom.  Terms of order $g^2$ in \LJV\ lead to
contact terms which are set such that the functional form in
\currentsmom\ is valid.  (A familiar example is the seagull term
in scalar electrodynamics.)

The effective Lagrangian for the gauge supermultiplet is
 \eqn\LeffV{\eqalign{ \delta\CL_{eff} &= \half
g^2\tilde C_0(0)D^2 -g^2\tilde
C_{1/2}(0)i\lambda\sigma^\mu\partial_\mu\bar\lambda -{1\over4}
g^2\tilde C_1(0)F_{\mu\nu}F^{\mu\nu}\cr
 &\qquad -{1\over2}g^2(M\tilde B_{1/2}(0)\lambda\lambda+c.c.)+ \dots
 }}
where the ellipses represent terms with  higher powers of
momentum. From this expression, we see that $\tilde C_a$
correspond to wavefunction renormalizations of $D$, $\lambda$ and
$A$, while $\tilde B_{1/2}$ corresponds to renormalization of the
gaugino mass. When supersymmetry is unbroken, the former must be
all the same and the latter must be zero, as stated in
\susyunbroken.

The divergent parts in $\tilde C_a$ \highmom\ clearly represent a
change in the beta function of the gauge fields, while the
supersymmetry breaking finite parts represent different thresholds
for these fields. To be more precise, we find that the
contribution to the beta function from the hidden sector fields is
given by \eqn\betac{ \Delta b = -(2\pi)^4 c } In other words,
 $b_{high}=b_{low}+\Delta b$, where the subscripts denote the
effective theories above and below the scale $M$ of the hidden
sector. Note that $\Delta b$ is always negative; this is expected
since the contribution to the beta function from charged matter is
always negative.\foot{As mentioned in the introduction, our
formalism does not include the case of ``gauge messenger" models
where the SM gauge group is nontrivially embedded into a larger
group such as $SU(5)_{GUT}$. In such models, the contribution of
the hidden sector to the beta function can have either sign.}

\ifig\neffrbfig{The graphical description of the contributions of
the two point functions to the soft masses.  (a) represents the
gaugino mass contribution from $\langle j_\alpha j_\beta\rangle$.
In (b)-(e) the various contributions to the soft scalar masses are
given: (b) $\langle J \rangle$, (c) $\langle JJ \rangle$, (d)
$\langle j_\alpha \bar j_{\dot\alpha}\rangle$, and (e) $\langle
j_\mu j_\nu \rangle$.  It should be stressed that the blobs in the
figures represent hidden sector correlation functions.  The
leading contribution in theories with messengers arises from one
loop of the messengers, but in general when there are no
messengers, it is more
complicated.}{\epsfxsize=1\hsize\epsfbox{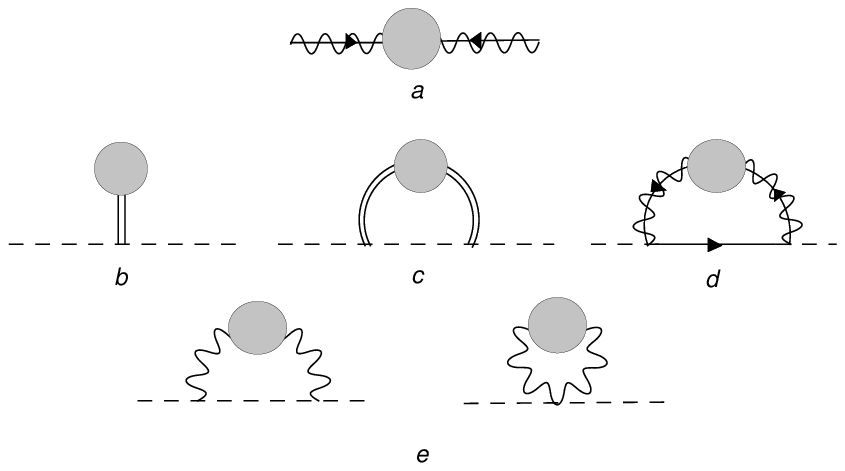}}

So far we have discussed the simpler case of a single $U(1)$ gauge
group here, in the case of the actual MSSM one has to consider the
separate $SU(3)$, $SU(2)$ and $U(1)$ gauge groups. We will label
the gauge groups by $r=3$, 2, 1, respectively. If we want the
gauge couplings to unify, then the value of $c^{(r)}=c$ must be
independent of $r$ (assuming $SU(5)$ normalization of the $U(1)$
factor of course) and we want the thresholds
$\tilde{C}_a^{(r)}(0)$ to depend weakly on $r$. Moreover, if we
want perturbative unification, then there is an upper bound on the
magnitude of $c$. These are examples of some completely general
constraints on the SUSY breaking sector that can be derived using
our formalism.

Now, it is straightforward to find the sfermion and gaugino masses
of the MSSM. In Figure 1 we show the diagrams involving the current
correlation functions which are responsible for the
MSSM soft masses.

The gaugino masses arise at tree level in the effective theory
\LeffV; to leading order they are given by
 \eqn\mgaugino{
 M_r = g_r^2M\tilde B_{1/2}^{(r)}(0).
 }
(Starting from this equation we use the hypercharge normalization
of $g_1$ which differs from the GUT normalization by $\sqrt
{5/3}$.)

We now compute the sfermion masses.  When $\langle J\rangle$ is
nonzero we get a tree level contribution to the sfermion $\tilde
f$ (the superpartner of the fermion $f$) mass squared of the form
$ g_1^2Y_{f}\zeta$ where $Y_{f}$ is the $U(1)$ hypercharge of the
sfermion.  A more interesting effect arises at one loop. As we
will soon see, the typical momentum in the loop is of order $M$,
and therefore the low momentum effective Lagrangian \LeffV\ cannot
be used. Instead, we use the full momentum dependence in the
correlators \currentsmom\ in three different one-loop diagrams,
one with an intermediate $D$, one with an intermediate $\lambda$
and one with an intermediate $V$. We easily find
 \eqn\sfermions{
 m_{\tilde{f}}^2=g_1^2Y_{f}\zeta + \sum_{r=1}^3 g_r^4\,c_2(f;r)
 A_r}
where $c_2(f;r)$ is the quadratic Casimir of the representation of
$ f$ under the $r$ gauge group; and
 \eqn\Ardef{\eqalign{
  A_r &\equiv -\int {d^4p\over (2\pi)^4}{1\over p^2}\Big(3\tilde
 C_1^{(r)}(p^2/M^2)-4\tilde C_{1/2}^{(r)}(p^2/M^2)+\tilde
 C_0^{(r)}(p^2/M^2)\Big) \cr
  &=  -{M^2\over 16\pi^2}\int dy\,\Big(3\tilde C_1^{(r)}(y)
  -4\tilde C_{1/2}^{(r)}(y)+\tilde C_0^{(r)}(y)\Big)
  }}
As stated above, the typical momentum in \sfermions\ is of order
$M$ rather than zero.

Although we did not prove it in general, the integrals in \Ardef\
should be UV convergent.  Otherwise we would need a counter term
for the sfermion masses which cannot be present in a theory with
spontaneously broken supersymmetry.

We make two comments about these results. First, it is clear
from this formalism that the gaugino masses are not a priori
related to the sfermion masses, nor are they necessarily related
to the change in the beta function \betac. Thus, there is no a
priori reason why one cannot have gauge coupling unification
without gaugino mass unification in general models of gauge
mediation. Second, we see from
\sfermions\ the well-known fact that an effective FI term $\zeta$
can be quite dangerous for gauge mediation, since it leads to a
non-positive definite (i.e.\ proportional to hypercharge)
contribution to the sfermion masses at lower-order in the gauge
couplings. Thus to avoid tachyonic slepton masses, usually it is
assumed that some symmetry forbids $\zeta$ (see e.g.\ the
``messenger parity" of \dimgiud). In our general formalism, we can
characterize this symmetry quite simply as an invariance of the
hidden sector under a ${\bf Z}_2$ symmetry which acts on $J$ as
$J\to -J$.

\newsec{Mass relations}

\subsec{Relations at the scale $M$}

In the previous section, we have seen how all the MSSM sfermion
masses are completely determined in terms of four real numbers
$(\zeta,\,A_1,\,A_2,\,A_3)$ which are derived from correlation
functions in the SUSY-breaking sector. In this section, we analyze
how this general result constrains the MSSM spectrum and leads to
definite relations among the sfermion masses.  We first consider
the commonly assumed case $\zeta=0$. Then there must be two
relations amongst the sfermion soft masses which are valid in
general. These mass relations can be easily derived by using the
facts that each generation of the MSSM is separately anomaly free
in $U(1)_Y$, and the mixed $U(1)_{B-L}$ -- gauge anomalies also
vanish.  From the general form of the sfermion masses \sfermions,
it follows that \eqn\mrels{ \Tr\, Y m^2 = \Tr\, (B-L)m^2 = 0 }
where the trace is over the MSSM sfermions in a given generation,
and $Y$ and $B-L$ stand for the hypercharge and $U(1)_{B-L}$
quantum numbers of the given sfermion, respectively. More
explicitly, the mass relations are given by
 \eqn\massrell{\eqalign{
 & m_{Q}^2-2 m_{U}^2+m_{D}^2-m_{L}^2+m_{E}^2 = 0 \cr
 & 2m_Q^2-m_U^2-m_D^2-2m_L^2+m_E^2=0\cr }}
These relations have been derived before in the context of various
specific SUSY-breaking models (see e.g.\
\refs{\pierre\faraggi\kawamura\MartinZB-\spectroscopy}). More
recently, they have been discussed in \hiddenren\ in the context
of models with strong hidden sector renormalization effects.
However, as our discussion makes clear, these relations are
completely general features of gauge mediation, which do not
depend on any specific form of the hidden or messenger sector
(indeed, there need not even be any invariant distinction between
the two). Thus, these relations offer a completely model
independent test of gauge mediation which could in principle be
carried out at the LHC or the ILC. Moreover, these sum rules could
in principle be used to distinguish gauge mediation from other
popular mediation schemes. In particular, the $B-L$ mass relation
is violated in mSUGRA and various modifications of anomaly
mediation which fix the slepton mass problem.

Next let us discuss the case that $\zeta\ne 0$. Then there
should only be one relation amongst the sfermion soft masses.
Indeed, it follows immediately from \sfermions\ that
\eqn\mrelsii{\eqalign{
 & \Tr\, Y m^2 -g_1^2\zeta \Tr Y^2 = 0\cr
 & \Tr\, (B-L)m^2  -g_1^2\zeta \Tr\, (B-L)Y = 0
 }}
which means that the linear combination $\Tr\,Y m^2 - {5\over 4}
\Tr\,(B-L)m^2 =0$ defines the one surviving mass relation.
Explicitly, this takes the form
 \eqn\included{
  6 m_{Q}^2-9 m_{D}^2+3m_{U}^2-6 m_{L}^2+m_{E}^2 = 0.
 }
Again, we emphasize that this sum rule is a completely model
independent prediction of gauge mediation.

Finally, let us point out that there would be in principle two
more relations relating the Higgs soft masses to the each other
and the sfermion masses.  However, since the Higgs soft masses
generally pick up an additional contribution from whatever effect
which generates $\mu$ and $B\mu$ (see e.g.\ section 5), we expect
that these additional mass relations will in general not be robust
predictions of general gauge mediation. The same statement might
apply also to the third generation mass relations, since the top
Yukawa is large.  On the other hand, since the two light
generations couple to the Higgs fields very weakly, the precise
details of the mechanism which generates $\mu$ and $B\mu$ hardly
affect these mass relations.

\subsec{Corrections to the sum rules}

The sum rules derived in Section 4.1 hold at the characteristic
mass scale $M$, which on general grounds must be at or above the
electroweak scale (in models with messengers $M$ can be thought of
as the messenger scale).   In general one must take into account
the running of the soft masses in the MSSM in order to obtain
low-energy spectrum, and this could potentially affect the sum
rules. In fact, we will see that the sum rules for the first and
second generation are quite robust under MSSM RG evolution. For
simplicity, we will assume $\zeta=0$ in this subsection.

Defining $S_Y^{(i)}=\Tr\, Y m_i^2$ and $S_{B-L}^{(i)}=\Tr\,
(B-L)m_i^2$ as the sum rules for the $i$th generation and
$S=\sum_i S_Y^{(i)}+m_{H_u}^2-m_{H_d}^2$, it is straightforward to
compute for the first two generations, using e.g.\ the formulas in
\MartinNS, the one-loop running of $S_Y^{(i)}$ in the MSSM:
 \eqn\diffS{
16\pi^2 {dS_Y^{(i)}\over dt} = 2 g_1^2 (\Tr\, Y^2) S } and
 \eqn\diffSBL{
 16\pi^2 {dS_{B-L}^{(i)}\over dt} = 2 g_1^2 (\Tr\, Y(B-L)) S }
where, again, the trace runs over just one  sfermion generation.
(For the third generation there are additional complications due
to the Yukawa couplings and $A$ terms.) In gauge mediation defined
without any modification in the Higgs sector, $S=0$ at $M$, and so
these sum rules are preserved at all scales. However since we are
allowing for potential modification to the Higgs sector in Section
5, we should keep in mind that there is in general an
inhomogeneous correction piece $m_{H_u}^2-m_{H_d}^2$ for both
$S_Y^{(i)}$ and $S_{B-L}^{(i)}$. Fortunately, since these
corrections are proportional to $\alpha_1$, they are typically
small for reasonable values of $m_{H_u}^2$ and $m_{H_d}^2$.
Additionally there are also small corrections due to the MSSM
D-terms after EWSB which can be found in \MartinZB\ and are $\CO
(m_z^2)$.

\newsec{Comments on the $\mu$/$B\mu$ problem}

One of the standard difficulties in gauge mediation models is the
$\mu/B\mu$ problem: how to generate the couplings
 \eqn\muBmu{
 B\mu H_uH_d + \int d^2 \theta \mu H_u H_d + c.c.
  }
with $\mu$ and $B$ of the right order of magnitude. Let us try to
address this in our very general framework.

Clearly, we need to couple the two Higgs fields $H_{u,d}$ to the
hidden sector.  One approach is to assume the existence of a
hidden sector chiral operator $\CA$ with coupling
 \eqn\CAcoupling{\lambda\int d^2 \theta \CA H_uH_d }
and vev
 \eqn\CAvev{ \lambda\langle \CA\rangle = \mu + \theta^2 B\mu .}
The operator $\CA$ can be a fundamental field in the hidden sector
theory. For example, this is the case in the NMSSM, if we view the
singlet field of that model as a part of the hidden sector and
identify it with $\CA$. A problem with that is that unless certain
discrete symmetries are imposed, a large tadpole for $\CA$ is
generated, leading to a need for fine tuning \polchinski.
Alternatively, $\CA$ can be a hidden sector composite field whose
short distance dimension is $\Delta>1$.  Its expectation value is
naturally of order $M^\Delta+ \theta^2 M^{\Delta+1}$. However, in
this case the coupling $\lambda$ in \CAcoupling\ is dimensionful and
it is suppressed by a power of a large scale, e.g. $\lambda \sim
1/M_{Planck}^{\Delta-1}$. Therefore, the effect of the interaction
\CAcoupling\ is negligible.  Similar comments apply to other
couplings like $\int d^4 \theta \CA^\dagger H_uH_d$.

The only kinds of couplings for which these comments do not apply
are
 \eqn\HuHdcoupling{\int d^2 \theta (\lambda_u \CO_uH_u + \lambda_d
 \CO_dH_d )}
where $\CO_{u,d}$ are composite operators with appropriate
$SU(2)_L\times U(1)_Y$ quantum numbers and $\lambda_{u,d}$ are
coupling constants.  For example, such operators can originate at
short distance from bilinears in the charged hidden sector fields.
In this case the coupling constants $\lambda_{u,d}$ are
dimensionless.  Examples of such couplings appear in models with
messengers, see e.g.\ \refs{\dgp}. In the coming discussion we
will assume that the short distance dimension of $\CO_{u,d}$ is two.

Above we defined gauge mediation as a situation in which in the
limit $\alpha_{1,2,3} \to 0$ the theory decouples into a hidden
sector and an MSSM sector.  In the presence of the couplings
\HuHdcoupling\ we must extend this definition to include the limit
$\lambda_{u,d} \to 0$.  Then, just as we have used perturbation
theory in $\alpha_{1,2,3} $ to examine the effect of the hidden
sector on the MSSM, we can also expand in $\lambda_{u,d}$.  At
leading order only couplings of the Higgs fields are affected.
Using the hidden sector correlation functions
 \eqn\hiddenO{\langle \CO_u \CO_u^\dagger \rangle \qquad ; \qquad
 \langle \CO_d \CO_d^\dagger \rangle \qquad ; \qquad
 \langle \CO_u \CO_d \rangle }
where $\CO_{u,d}$ stand for the full chiral superfields, we can
generate Higgs masses, and the couplings \muBmu. (In hidden sector
models with multiple scales, we can also generate operators of the
form \CAcoupling, etc.\ with $\CA$ given by a composite hidden
sector operator. But in these cases the operator will be
suppressed by a high scale in the hidden sector, not
$M_{Planck}$.)  Assuming that the correlation functions in
\hiddenO\ are given by powers of $M$, we naturally find $\mu \sim
\lambda_u\lambda_d M$ and $B \sim M$. For $\lambda_u\lambda_d \sim
{\alpha \over 4\pi}$ the generated $\mu$ is of the right order of
magnitude, but $B$ is too large. This is a well known problem with
gauge mediation models (see e.g. \dgp).  In this general language,
we see that the problem is clearly in the assumption that all the
correlation functions are given by powers of $M$. One can
certainly imagine that with the right hidden sector, some of the
correlation functions in \hiddenO\ are smaller than others, and
this could lead to $B \roughly< \mu $ which are of the same order
as the other soft breaking terms. For instance, this could
conceivably arise either from anomalous dimensions in the hidden
sector theory or from an approximate symmetry.\foot{The use of
anomalous dimensions for the $\mu/B\mu$ problem was recently
discussed in \refs{\martinmu,\hidrentwo}. These models have two
scales, with messengers being integrated out at the higher scale.}
The former possibility is not inconceivable, especially if the
hidden sector is strongly coupled, since the dimensions of the
operators $\CO_{u,d}$ (unlike the currents discussed above) are
not necessarily protected by any symmetry. Finally, we point out
that using higher point functions in the hidden sector, we can
generate the effective dimension five and six operators of
\DineXI\ thus potentially avoiding the little hierarchy
problem\foot{These operators had been noticed and analyzed by
various authors before \DineXI, see e.g.\ \BrignoleCM. Our
analysis here is in the spirit of \DineXI\ which considered a low
energy effective Lagrangian obtained by integrating out generic
short distance theories.  Then these operators are the dominant
ones in a systematic expansion.}.

It is clear that a much more detailed analysis of the correlation
functions \hiddenO\ is needed before we can conclude whether a
typical hidden sector model can lead to a fully satisfactory
solution of all phenomenological problems.  We intend to return to
such an analysis in the near future.

\bigskip

\noindent {\bf Acknowledgments:}

We would like to thank N.~Arkani-Hamed, M.~Dine, Y.~Nakayama, A.~Nelson,
S.~Thomas and T.~Volansky for useful discussions. The work of PM
and NS was supported in part by DOE grant DE-FG02-90ER40542. The
work of DS was supported in part by NSF grant PHY-0503584. Any
opinions, findings, and conclusions or recommendations expressed
in this material are those of the author(s) and do not necessarily
reflect the views of the National Science Foundation.

\appendix{A}{A Simple Example}

To illustrate the general techniques presented in the text, let us
consider the simple case of minimal gauge mediation (for a $U(1)$
toy model), where
 \eqn\chvexiii{
\delta \CL = \int d^4\theta \,\left(\phi^\dagger e^{2gV}
\phi + \tilde\phi^\dagger e^{-2gV}\tilde\phi\right)
+ \left( \int d^2\theta\,\lambda X \phi\tilde\phi + c.c.\right)
 }
with $\langle X\rangle = M+\theta^2F$. (Without loss of
generality, we will take $M$ and $F$ to be real and set
$\lambda=1$.) Then we have complex scalar fields
$\phi_\pm=(\phi\pm\tilde\phi^*)/\sqrt{2}$ with masses
$m_\pm^2=M^2\pm F$; and two fermions $\psi$,
$\tilde\psi$ which both have mass $m_0=M$. The components
of the current superfield are
 \eqn\Jcompexiii{\eqalign{
 J(x) &= \phi^*\phi(x)-\tilde\phi^*\tilde\phi(x)\cr
 j(x) &= -\sqrt{2}i(\phi^*\psi(x)-\tilde\phi^*\tilde\psi(x))\cr
 \bar j(x) &= \sqrt{2}i(\phi\bar\psi(x)-\tilde\phi\bar{\tilde\psi}(x))\cr
  j_\mu(x)& = i(\phi\partial_\mu\phi^*(x)-\phi^*
  \partial_\mu\phi(x)-\tilde\phi\partial_\mu
  \tilde\phi^*(x)+\tilde\phi^*\partial_\mu\tilde\phi(x)) +
  \psi\sigma_\mu\bar\psi(x) - \tilde\psi\sigma_\mu\bar{\tilde\psi}(x)
 }}
{}From this, we obtain the correlators:
 \eqn\correxiii{\eqalign{
 \langle J(0)\rangle &= 0 \cr
    \langle J(x)J(0)\rangle &=
  2D(x;m_+)D(x;m_-)\cr
    \langle j_\alpha(x)\bar j_{\dot\alpha}(0)\rangle &=
  -2i(D(x;m_+)+D(x;m_-))\sigma_{\alpha\dot\alpha}^\mu\partial_\mu D(x;m_0)\cr
  \langle j_\mu(x)j_\nu(0)\rangle &=
  2\Bigg(\Big(\partial_\mu D(x;m_+)\partial_\nu D(x;m_+)
  -D(x;m_+)\partial_\mu\partial_\nu D(x;m_+)\Big)\cr
  &\quad +\Big(\partial_\mu D(x;m_-)\partial_\nu D(x;m_-)
  -D(x;m_-)\partial_\mu\partial_\nu D(x;m_-)\Big)\cr
&\quad +  2\eta_{\mu\nu}\Big(\partial^\rho D(x;m_0)\partial_\rho
D(x;m_0)-m_0^2D(x;m_0)^2)\Big)  -4\partial_\mu D(x;m_0)\partial_\nu
D(x;m_0) \Bigg)\cr
 \langle j_\alpha(x)j_\beta(0)\rangle &=
  -2(D(x;m_+)-D(x;m_-))\epsilon_{\alpha\beta}m_0D(x;m_0)\cr
     }}
where
 \eqn\Dgen{
 D(x;m)=\int {d^4p\over (2\pi)^4}\,{ie^{ipx}\over p^2-m^2}
 }
is the propagator for a scalar field with mass $m$. The
expressions in \correxiii\ were derived by performing the
free-field contractions on the correlators. Note that they are
only valid for $x\ne 0$; for $x\to 0$ one must be more careful
about including delta-function contact terms necessary for current
conservation. These are most easily determined in momentum space,
and we will take them into account below. As a check of these
expressions, note that they satisfy the relations \susyunbroken\
in the SUSY limit $m_+=m_-=m_0$, with $C_a=2D(x;m_0)^2$.

{}From the correlators, we can extract the functions $\tilde C_a$,
$\tilde B_{1/2}$ by Fourier transforming the RHS of \correxiii,
substituting \Dgen, and comparing with \currentsmom. For example,
the first correlator of \correxiii\ yields
 \eqn\tildeCzeroexiii{\eqalign{ \tilde C_0 = \int
d^4x\,e^{ipx}\langle J(x)J(0)\rangle & = \int d^4x\, e^{ipx}\Big(
2D(x;m_+)D(x;m_-)\Big)\cr
  &=2\int {d^4q\over (2\pi)^4} {1\over (q^2+m_+^2)((p+q)^2+m_-^2)} \cr
  }}
Performing similar manipulations for the other correlators, we
obtain the final expressions
 \eqn\tildeCexiii{\eqalign{
&\tilde C_0 = 2\int {d^4q\over (2\pi)^4} {1\over (q^2+m_+^2)
((p+q)^2+m_-^2)}\cr
 &\tilde C_{1/2} = -{2\over p^2}\int {d^4q\over (2\pi)^4}
  \left({1\over (p+q)^2+m_+^2}+{1\over (p+q)^2+m_-^2}\right)
  {p\cdot q\over q^2+m_0^2}\cr
 &\tilde C_1 = -{2\over 3p^2}\int{d^4q\over (2\pi)^4}\Bigg(
 {(p+q)\cdot(p+2q)\over (q^2+m_+^2)((p+q)^2+m_+^2)}+(m_+\to m_-)\cr
 &\qquad \qquad\qquad +{4q\cdot(p+q)+8m_0^2 \over
 (q^2+m_0^2)((p+q)^2+m_0^2)}
 -{4\over q^2+m_+^2}-{4\over q^2+m_-^2}\Bigg)
}} and \eqn\tildeBexiii{ M\tilde B_{1/2} = 2m_0\int {d^4q\over
(2\pi)^4}\left( {1\over q^2+m_-^2}-{1\over
 q^2+m_+^2}\right){1\over (p+q)^2+m_0^2} }
In $\tilde C_1$, we have included contributions from contact terms
(the last two terms in the last line of \tildeCexiii). These are
required in order for $\langle j_\mu(p)j_\nu(-p)\rangle $ to
satisfy the Ward identity as in \currentsmom.

At large $p$, the functions $\tilde C_a$ all have the form
 \eqn\largepexiii{\eqalign{
\tilde C_a &= {1\over8\pi^2}\left(\log{\Lambda^2\over p^2}+1
\right) + {1\over 8\pi^2 p^2}\left(m_-^2\log {m_-^2\over
p^2}+m_+^2\log {m_+^2\over p^2}-m_-^2-m_+^2\right) \cr
 &\qquad\qquad\qquad\qquad\qquad\qquad\qquad\qquad
  +  \CO(1/p^4,\,(\log p^2)/p^4)
 }}
i.e.\ they all agree up to $\CO(1/p^2)$ but not necessarily at
$\CO(1/p^4)$.  Note that the agreement at $\CO(1/p^2)$ depends on
the fact that the messengers satisfy the supertrace relation,
$m_-^2+m_+^2=2m_0^2$. In general, we expect that the functions
$\tilde C_a$ should agree up to $\CO(1/p^2)$ if supersymmetry is
spontaneously broken. One important consequence of this is that
the integral \Ardef\ for the sfermion masses is always UV
finite, even though the individual terms contributing to it are
not.

Finally, let us compare to the well-known formulas for the
one-loop  gaugino and two-loop sfermion masses of minimal gauge
mediation. From \mgaugino\ and \tildeBexiii, we find
 \eqn\mgauginoexiii{
M_{\lambda} = 2g^2m_0 \int {d^4q\over (2\pi)^4}\left({1\over
q^2+m_-^2}-{1\over q^2+m_+^2}\right){1\over q^2+m_0^2}=
 {\alpha\over 4\pi}{F\over M}\times 2g(x)
}
where $x=F/M^2$ and
\eqn\gofxdef{
g(x)= {(1-x)\log(1-x)+(1+x)\log(1+x)\over x^2}
 }
This agrees precisely with the answer in \MartinZB; the factor of
two in \mgauginoexiii\ is the Dynkin index for the pair of
messengers; more generally it would be $2Y^2$ where $Y$ is the
$U(1)$ charge.

Next we compare with the formula for the sfermion masses
\MartinZB:
 \eqn\massesmgm{\eqalign{
 m_{\tilde f}^2  &= g^4(A_{0}+A_{1/2}+A_1)
}}
where
\eqn\Adef{\eqalign{
 & A_0= -2G_2(m_+,m_-)\cr
 & A_{1/2}= 4G_1(0)(G_0(m_+)+G_0(m_-)-2G_0(m_0))+4G_2(m_+,
 m_0)+4G_2(m_-,m_0)\cr
  &\qquad\qquad +4(m_+^2-m_0^2)G_3(m_+,m_0)+4(m_-^2-m_0^2)
  G_3(m_-,m_0)\cr
 & A_1= -4G_1(0)
 (G_0(m_+)+G_0(m_-)-2G_0(m_0)) - G_2(m_+,m_+)-G_2(m_-,m_-)\cr
 &\qquad -4G_2(m_0,m_0)-4m_+^2G_3(m_+,m_+)-4m_-^2G_2(m_-,m_-)
 +8m_0^2G_3(m_0,m_0)
}}
and
\eqn\Gdefs{\eqalign{
 & G_0(m) = \int {d^4p\over(2\pi)^4}{1\over p^2+m^2}\cr
 & G_1(m) = \int {d^4p\over (2\pi)^4} {1\over (p^2+m^2)^2}\cr
 & G_2(m_1,m_2)\equiv \int {d^4p\over (2\pi)^4}{1\over p^2}
 \int {d^4q\over (2\pi)^4}{1\over (q^2+m_1^2)((p+q)^2+m_2^2)}\cr
  & G_3(m_1,m_2)\equiv \int {d^4p\over (2\pi)^4}{1\over p^4}
  \int {d^4q\over (2\pi)^4}{1\over (q^2+m_1^2)((p+q)^2+m_2^2)}
}
 }
(Despite appearances, these functions are symmetric under
interchange of $m_1$ and $m_2$.) Comparing with \sfermions\ and
\tildeCexiii, we find that the individual $A_a$ agree precisely
with the contributions from $\tilde C_a$, respectively. Note that
the contact terms included in \tildeCexiii\ were crucial for the
agreement between $A_1$ and the contribution from $\tilde C_1$.

\listrefs
\end